\documentclass[epj]{svjour}
\usepackage{epsfig}

\def\be{\begin{equation}}
\def\ee{\end{equation}}
\def\ba{\begin{eqnarray}}
\def\ea{\end{eqnarray}}
\def\br{\begin{array}}
\def\er{\end{array}}

\begin{document}
\title{Self-Quenched Dynamics}

\author{J\'anos T\"or\"ok\inst{1}\thanks{e-mail:
torok@planck.phy.bme.hu}, Supriya Krishnamurthy\inst{2},
J\'anos Kert\'esz\inst{1} and St\'ephane Roux\inst{3}}
\institute{$^1$ Department of Theoretical Physics, 
Institute of Physics, Budapest University of Technology, 
Budafoki \'ut 8, Budapest, H-1111, Hungary \\
$^2$ Department of Theoretical Physics, University of Oxford
1 Keble Road, Oxford OX1 3NP, UK \\
$^3$ Surface du Verre et Interfaces, UMR CNRS/Saint-Gobain,
39 Quai Lucien Lefranc, 93303 Aubervilliers Cedex, France
}

\date{\today}

\abstract{We introduce a model for the slow relaxation of an energy
landscape caused by its local interaction with a random walker whose
motion is dictated by the landscape itself. By choosing relevant
measures of time and potential this {\it self-quenched} dynamics can
be mapped on to the ``True'' Self-Avoiding Walk model. This
correspondence reveals that the average distance of the walker at time
$t$ from its starting point is $R(t) \sim {\log(t)}^{\gamma}$, where
$\gamma=2/3$ for one dimension and $1/2$ for all higher dimensions.
Furthermore, the evolution of the landscape is similar to that in
growth models with extremal dynamics.}

\PACS{{05.40.Fb}{Random walks and Levy flights} \and
{05.65.+b}{Self-organized systems} \and
{05.50.+q}{Lattice theory and statistics} }
\maketitle

\section{Introduction}
The motion of random walkers in a random environment is one of the
basic problems in the physics of disordered systems
\cite{Alexander,Sinai,Fisher}. It is known that the effect of the
environment on the walker results in anomalous diffusion in some cases
and logarithmically slow diffusion in others \cite{Alexander,Sinai}.
Apart from their intrinsic interest, these simple models also find
applications in several physical processes, such as the diffusion of
electrons in a disordered medium \cite{Bernasconi} or glassy activated
dynamics \cite{Fisher}. In particular, the Sinai model \cite{Sinai}
has been extensively studied from this point of view. It is known
that for this model, the walker becomes logarithmically slow, moving
as $R(t) \sim {{\log}^{2}(t)}$ where $R(t)$ is the average distance at
time $t$ of a walker from the launching point. Apart from this,
two-time aging dynamics studied in this model also provide close
analogies to glassy phenomenology \cite{Fisher}. 

In another framework, a ``trapping'' model (\cite{monthus} and
references therein) was introduced to provide a simple example of the
glass transition. A single walker explores a random landscape of
energy traps with e.g. an exponential distribution, from which it can
escape through activated hops. At low temperatures, it was shown that
the system cannot reach a steady state due to the unfavorable
competition between the depth of the visited traps and the time needed
to escape from them. The corresponding slow dynamics, and
disappearance of an equilibrium state was interpreted as a glass
transition. In this mean-field model, introducing an interaction
between the walker and the random energy landscape was shown to not
change the results \cite{monthus}. However, the fact that the trapping
time distribution is a power-law provides a natural (statistical)
history dependence. (The dynamics is controlled by the deepest energy
well visited so far). In the model we introduce here, the walker
modifies its environment; aging and slow dynamics result purely
because of this interaction. 

Active walker models where the walker and the environment mutually
affect each other have been studied earlier in different contexts.
One such model --- the Eulerian Walkers Model (EWM) --- has been
studied \cite{Eulwalk} within the framework of self-organized
criticality (SOC) \cite{Bak}. The ``landscape'' here is defined by an
arrow at each site. The walker follows the direction of the arrow at
its site after which the direction of the arrow is changed according
to some fixed rules. Besides many correspondences between the EWM and
the Abelian Sandpile Model of SOC \cite{ddhar}, it was also shown that
the motion of the walker was sub-diffusive in two dimensions, {\it
i.e.} $R(t) \sim t^{1/3}$. In one dimension, $R \sim t^{1/2}$ due to a
very simple organization of the landscape under the rules. For $d>2$
it was argued that the walker is diffusive. Models of mutually
interacting walkers and landscape have also been studied extensively
in the context of pattern formation and biophysical applications, where
the emphasis is on the patterning of the medium under the influence of
multiple walkers \cite{activ_w,Czirok}. 

The Self-Avoiding Walk (SAW)  \cite{deGennes}, where the walker
is obliged to avoid its former path, and its modifications, can also
be regarded as active walks in the sense that the path of the walker in
these models is influenced by the trace it has  made. For later purposes
it is useful to mention here the so called ``True'' Self-Avoiding Walk
(TSAW) where the walker's probability to go to an already visited site
is a strongly decaying function of the number of visits to that
particular site \cite{TSAWorig}.

The aim of this paper is to study a simple model of an active walker
which exhibits {\it logarithmically} slow dynamics entirely due to the
local interaction of the walker with a self-created random environment
-- a kind of self-organized trapping or self-burying effect. Besides
the above mentioned connections with the physics of glasses, the model
in one dimension may also be regarded as a very simplified version of
a recently introduced model for the slow dynamics of sheared granular
media, where the wandering of shear bands and the related
restructuring of the material was shown to lead to extremely slow
relaxation processes and inhomogeneous aging \cite{TKKR}. 

\section{Definition of the SQW model}
Our model is defined as follows. A walker can move on a hyper-cubic
lattice in a $d$ dimensional space with linear size $L$ and $N = L^d$
number of sites. Periodic boundary conditions are imposed. A variable
$s_i$ chosen from a uniform random distribution $[0,1]$ is initially
assigned to every site $i$ on the lattice. We call the site with the
walker the {\it active} site.

The only permitted elementary moves for the walker are to the nearest
neighbours. At every time step a new random number $s_i(t)$ (uniformly
sampled between $0$ and $1$) is assigned to the active site $i$. If
this random number is larger than the value of the $s$ variable of
{\it all} the nearest neighbours, then the same site remains active.
Otherwise, the activity moves to the neighbouring site with the
largest value of $s$ and the same procedure is repeated. We have
chosen here a uniform distribution for $s$, however, this specific
form can be shown to play no role in the time evolution of the active
site. In the following we will refer to this model as a Self-Quenched
Walk (SQW).

As time goes on, the average value of $s$ decreases and as a result,
the probability that the activity moves to one of the neighbours
decreases. The definition of the model is thus quite inconvenient in
terms of numerical simulation since intervals when nothing happens
grow longer and longer with time. However, it is easy to circumvent
this difficulty. Let $\sigma$ be $s_{j}$, the largest $s$ value
amongst the neighboring sites of the active one, $i$. For the activity
to move to $j$, the $s_i$ value has to be smaller than $\sigma$ which
is an event with probability $\sigma$. Thus the waiting time before a
move is a Poisson process with a characteristic time $\tau=1/\sigma$.
Once $s_{i}<\sigma$, the evolution of the activity is deterministic.
Moreover, the distribution of $s_{i}$ when the activity moves to ${j}$
is uniform between 0 and $\sigma$. Therefore, we can directly
reproduce the evolution of the model in terms of the number of moves
$n$ rather than in time $t$. It turns out that the variable $n$ is
also more convenient for the analytical treatment.

Similarly, since the $s$ values quickly evolve towards 0, as $t$
increases, it is more convenient to use an equivalent parameterization
introducing $r \equiv -\log(s)$. The uniform distribution of $s$
between 0 and $\sigma$ implies that $r$ is distributed with a density
$e^{\rho-r}$ for $r>\rho=-\log(\sigma)$. Alternatively, we note that
$r-\rho$ is a random variable exponentially distributed from $0$ to
$\infty$. This reformulation allows simulations to be carried out
over practically unlimited times without loosing any accuracy.
Moreover, as we will see below, $r$ is the appropriate scale for
providing an accurate description of the long time regime.

Let us thus consider the motion of the activity (the walker) as a
function of the number of moves $n$, in a potential $V(i,n)$ [where
the value of the potential $ V(i,n)= r_i(n)$].
The walker moves to the neighbouring site with the {\it smaller} value
of $r$ [since $r \equiv -\log(s)$] after having changed the value of $V$
on the site it was on. The above statement can be made more 
quantitative in the following two coupled equations of evolution for
the walker and the $r$-landscape:
\ba
\label{Eq_c1}
\frac{d{\bf X}}{dn} &=& -\nabla V({\bf X}(n),n)+ {\bf \eta} (n) \\
\label{Eq_c2}
\frac{\partial V({\bf x},n)}{\partial n} &=&
\lambda \delta^{d} ({\bf x}- {\bf X}(n)) 
\ea
where $ \langle \eta (n) \rangle =0$ and $ \langle \eta (n) \eta 
(n^{\prime}) \rangle = C\delta (n-n^{\prime}) $. Here
${\bf X}(n)$ is the position of the walker at 'time' $n$.

Eq. \ref{Eq_c1} quantifies the rules of the SQW in any dimension. When
the walker is on a slope, it moves down towards the valley. When it is
in a flat region, it moves to any of the nearest neighbours with equal
probability (this is the reason for the uncorrelated noise term). Eq.
\ref{Eq_c2} accounts for the increase of the $r$-landscape at the
position of the walker. In this simplified continuum description, we
have neglected the randomness in the distribution of the local
increments in $r$, and only retained the average value $\lambda$.
However, as we shall see below, all the essential features of the
problem are contained in the equations. 

\section{Correspondence with the TSAW model}
It turns out that the Langevin equations (Eqs. \ref{Eq_c1} and
\ref{Eq_c2}) of the SQW in terms of $n$ and $r$ are the same as that
of the TSAW as a function of time and position \cite{TSAWorig}. The
definition of the TSAW model is the following: The walk takes place on
a $d$-dimensional hypercubic lattice. At any step the traveller may
move to any of the $2d$ nearest neighbours of the lattice site he is 
at. The probability of stepping to site $i$ depends on the number of
times $n_i$ this site has already been visited and is given by
\be
p_j=\exp(-gn_j)\left[\sum_{j=1}^{2d} \exp(-gn_j)\right]^{-1},
\ee
where the sum runs over all $2d$ nearest neighbours of the current
position of the walker, and $g$ is a positive parameter which measures
the intensity with which the walk avoids itself. Note that
the sum over $i$ of $p_i$ is equal to $1$, meaning that the traveller
never stays at the same site. This is similar to the
SQW when time is incremented in units of   $n$.

In the SQW the value of $s$ decreases exponentially on average since
each time the walker visits the site we multiply the $s$ value by a
random number taken from a uniform distribution between $0$ and $1$.
The value of $r=-\log(s)$ thus increases linearly with the number of
times the traveller has visited this site.

Further, in the SQW, whenever the walker can move, it goes 
deterministically to
the neighbouring site with the higher value of the potential. 
This is realized by the
$g\to\infty$ limit of the TSAW model.

The above mapping thus ensures that in the continuum limit the two
models are governed by the same Langevin equations. 
The role of the preassigned probability (Eqn. 3) in
the TSAW is played by the self-organized evolution of the walker and the
potential.

The TSAW has been studied exhaustively by means of
numerical simulations \cite{JCAdA,BerPie}, Flory theory
\cite{FamDaoud}, scaling analysis \cite{Pietronero} and later by exact
calculations \cite{BToth}. 

\begin{figure}
\centerline{\epsfxsize=0.8\hsize \epsffile{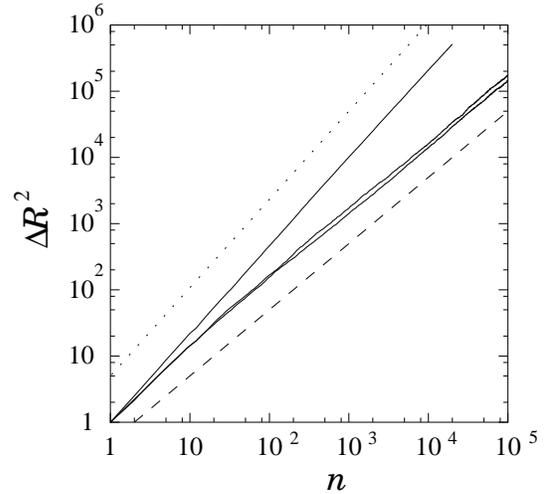}}
\caption{\label{Fig_rms}
The mean square distance from the origin covered by the walker in the
steady state, as a function of the number of moves $n$. The dotted
line shows power law with exponent $4/3$ (upper) and the dashed with
an exponent $1$ (lower). The curves are for dimensions $1$, $2$ and
$3$ from top to bottom respectively.
}\end{figure}

The critical dimension of the TSAW problem is $d_c=2$ above which the
mean field solution applies and the traveller's asymptotic behavior is
not influenced by the interaction with its former path and performs
basically a Brownian motion. Below two dimensions the trace of the
walker is a {\it fractional} Brownian motion. The root mean square
distance from the origin increases as
\be
\langle|{\bf X}(n) -{\bf X}(n')|^2\rangle^{1/2} \propto |n-n'|^\nu
\ee
with
\be
\nu=\cases{2/(d+2) & for $d\leq 2$ \cr
1/2 & for $d>2$}
\ee
Thus in one dimension the walker is super-diffusive with a Hurst
exponent of $\nu_{d=1}=2/3$ which is due to the repulsive interaction
with its former path. This asymptotic behavior is numerically verified
for the SQW in Fig. \ref{Fig_rms}.

We can define other related exponents through the scaling relations:
$X\rightarrow bX$, $n\rightarrow b^{z}n $ and $V\rightarrow
b^{\chi}V$, where $\chi$ is the so called {\it roughness exponent} of
the $r$ landscape and $z$ is the dynamic exponent. 
Equations \ref{Eq_c1} and \ref{Eq_c2} then predict
the following values for these exponents:

\be\br{llll}
\chi =1/2\quad &z=3/2 \quad &\textrm{ in\ } &d=1\\
\chi = 0&z=d &\textrm{ for\ } &d\geq 2.
\er\ee

We have confirmed the value of $\chi$ and $z$ by measuring the width
of the $r$ landscape in our model.

\section{Roughness of the potential}

The width of a self-affine
interface is defined as the root mean square fluctuation of the
interface from its mean value.
In a number of growth models, this width obeys the Family-Vicsek
scaling \cite{FS} with a dynamic exponent $z$ such that the overall
roughness follows
\be
w(n)\propto L^\chi\varphi\left({n\over L^z}\right)
\ee

\begin{figure}
\centerline{\epsfxsize=0.8\hsize \epsffile{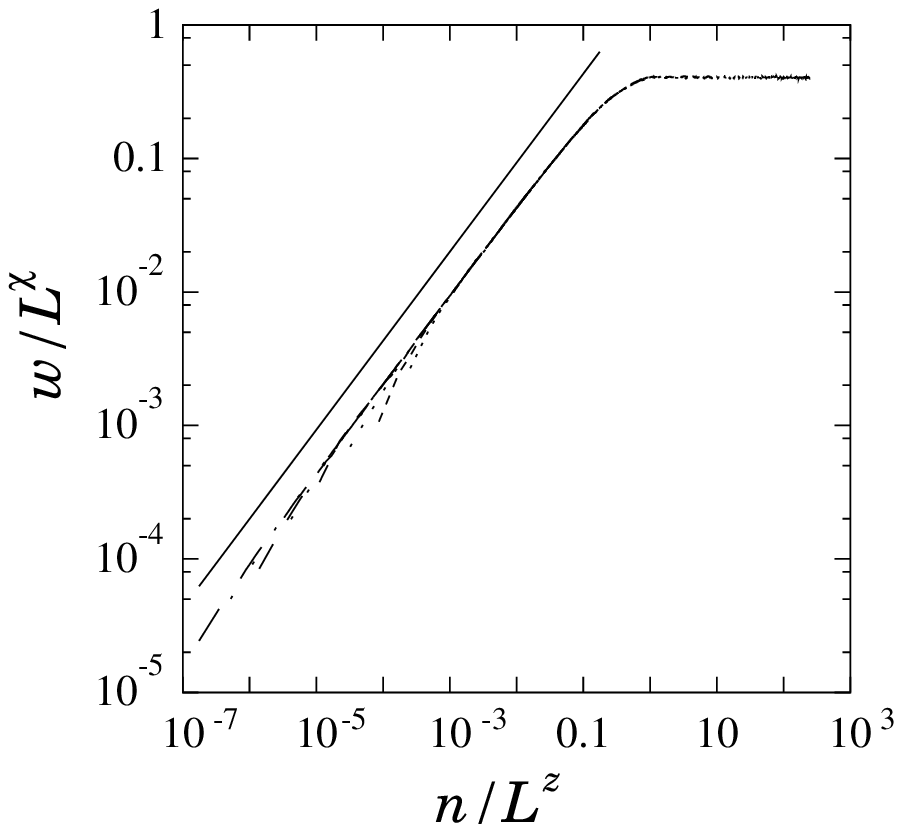}}
\centerline{\epsfxsize=0.8\hsize \epsffile{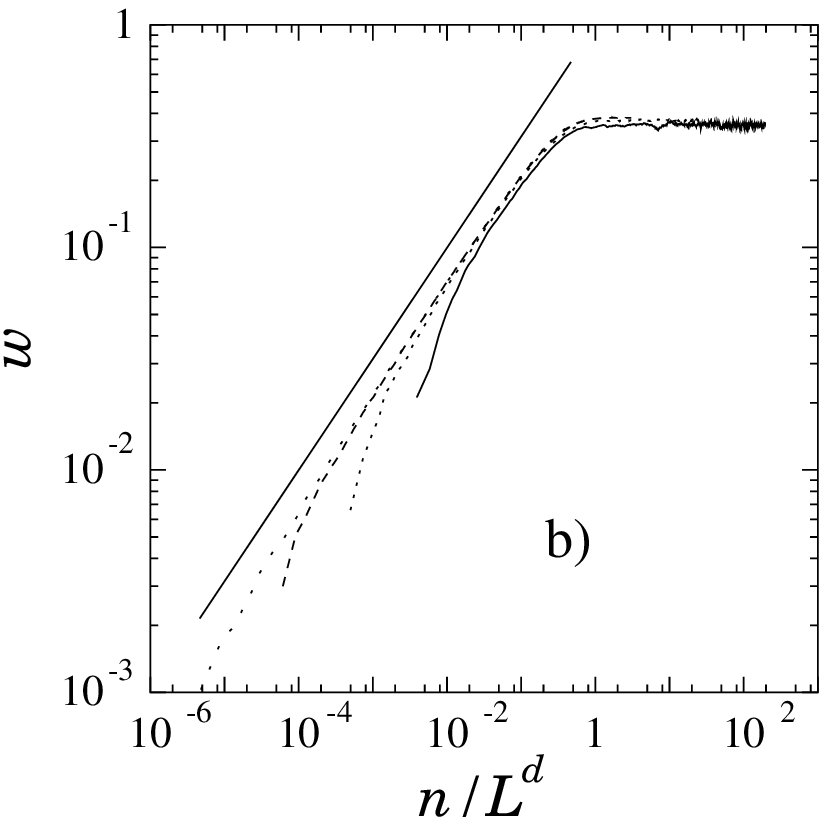}}
\caption{\label{Fig_width}
a) (Above) Average of squared roughness $w$ in one dimension scaled by
$L^{\chi}$ as a function of the number of moves $n$ scaled by the
dynamic exponent $L^z$. The different curves refer to system sizes
256*1,2,4,8,16; each averaged over 1000 realizations. The slope of the
solid line is $\beta=2/3$. \hfill\break
b) (Below) In 3 dimensions the steady state value of the width does
not depend on the system size. System sizes $8$, $16$, $32$, $60$ were
used each averaged over 1000 realizations. The slope of the solid line
is $\beta=1/2$.
} \end{figure}

Fig. \ref{Fig_width} shows our numerical determination of the width of
the $r$ landscape for five different system sizes in one and three
dimensions. The collapse with the above mentioned value of the
exponents indicates that the $r$ landscape is self-affine in one
dimension. However, the growth exponent $\beta$ describing the
roughening of the landscape fluctuations for early times ($n<n^* \sim
L^z$) as $w \sim n^\beta$ does not obey the Family-Vicsek scaling
$\beta = \chi/z$. It is instead given by
\be
\beta= (\chi + 1/2)/(1+\chi),
\ee
a formula typical for growth models with extremal dynamics
\cite{supri}. Thus our model, though it does not
contain a global extremum criterion, belongs to the class of extremal
growth models.

The reason why the interface growth (or the $r$ landscape evolution)
is similar to extremal dynamics is the following. The activity is a
'random' walk trapped by the maxima of the $r$-landscape. Before the
activity can escape, the landscape has to be filled. In order to
escape from a region of extent $\tilde l$, the number of moves to be
made is of the order of the size of the valley, $\tilde l^d$, times
its typical depth $\tilde l^\chi$. The growth is pinned everywhere
except in the immediate vicinity of the walker. The walker itself is,
however, in a hierarchically ordered valley structure with maxima of
increasing heights. Thus the interface progresses jerkily just as in
other extremal growth models. The relation $z= d+ \chi$ predicted by
Eqn. \ref{Eq_c2} is also known to occur in various extremal growth
models \cite{Maslov,Tang}.
   
The correspondence with extremal
dynamics does not hold above $d_c=2$. This is because above two
dimensions, the walker is no longer trapped by surrounding maxima. It
can also find its way around them instead of over them. As a result
the interface is no longer rough and $\chi =0$. However, though the
width saturates to a system size-independent value, the exponent
$\beta$ is non-zero for the following trivial reason. In the random
initial state, in $n$ steps, a number $n$ sites grow by, say, a
uniform amount $h$. Therefore the width $w$ scales trivially as $w
\sim hn^{1/2}L^{-1/2}$. The exponent $\beta$ is hence equal to $1/2$
for all $d\ge 2$.

\section{Real time behaviour}
We now turn to the behaviour of the walker in real time. In order to
understand this, we first note that the mean value of $r$ in the
steady state increases linearly with the number of moves. Hence the
mean increase of $r$ per site in the steady state is $1/N$, where
$N=L^d$ is the number of sites in the system.

\begin{figure}
\centerline{\epsfxsize=0.8\hsize \epsffile{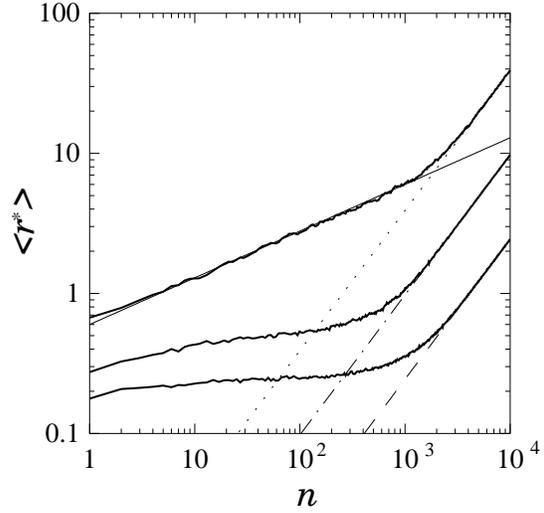}}
\caption{\label{Fig_alpha}
Average of $r^*$ as a function of the number of moves $n$, for
a system of size $L=256$, $L=32$ and $L=16$ in $1$, $2$ and $3$
dimensions respectively from top to bottom averaged over 1000 samples.
The solid line shows the a power-law of exponent $1/3$,
whereas the dashed lines are the asymptotic exact increase $r^*=n/N$.} 
\end{figure}

We now compute $\langle r^*(n)\rangle$ (the $r$ value of the active
site as a function of $n$) and observe two regimes that we describe by
the scaling assumption:
\be
\langle r^*(n)\rangle\propto n^\alpha\psi\left({n\over L^z}\right)
\ee
with $\psi(a)\propto a^0$ for $a\ll 1$ and $\psi(a)\propto a^{1-\alpha}$
for $a\gg 1$.
Therefore, for long times $\langle r^*(n)\rangle\propto
nL^{-z(1-\alpha)}$. However, in the long time regime, $r^*$ has to
increase at the same rate as the mean velocity of the front $\langle
r\rangle$, and 
hence, for $n\gg L^z$, $\langle r^*(n)\rangle =n/N=nL^{-d}$. This imposes
\be
z(1-\alpha)= d \quad{\rm or}\quad
\label{eqalpha}
\alpha=1-{d\over z}={\chi\over d+\chi}
\ee
i.e. $\alpha=1/3$ for $d=1$ and $0$ for $d \ge 2$ consistently with
our numerics as shown in Fig. \ref{Fig_alpha}.

\begin{figure}
\centerline{\epsfxsize=0.8\hsize \epsffile{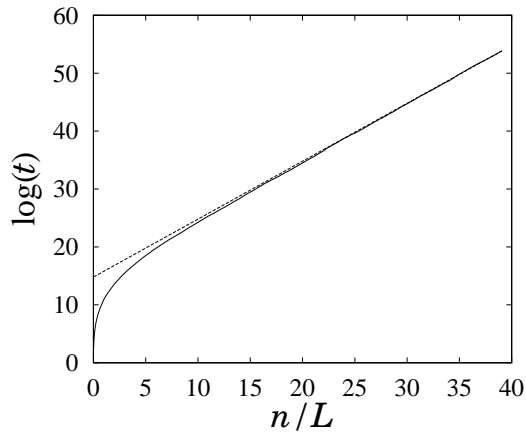}}
\caption{\label{Fig_realt}
Average of the logarithm of the time as a function of the
number of moves in $d=1$, scaled by the system size $n/L$. The
system size is $L=256$ and the average is performed over 1000 samples.
The dotted line is a line of slope 1, as theoretically expected.}
\end{figure}

Since, in the late stage regime, $r^*$ increases as $n/N$,
$s^*$ decreases as $e^{-n/N}$. Therefore the expectation value of the 
real time lapse between two consecutive moves $t(n)-t(n-1)$ is $1/s^*$ or
\be
t(n)-t(n-1)\approx e^{n/N}
\ee
\be
t(n)={e^{(n+1)/N}-1\over e^{1/N}-1}\sim Ne^{n/N}
\ee
This law refers in fact only to the mean value of $t$. However, the
distribution of each increment being exponentially distributed (a Poisson
process), the central-limit theorem applies, and the relative
standard deviation of $t$ with respect to its mean value vanishes.
Figure \ref{Fig_realt} shows the numerically determined time as a function of
the number of moves $n$.

The average value of $s$ as a function of time is thus 
\be
\langle s\rangle\propto \exp(-n/N)\propto \exp(-\log(t/N))\propto N/t
\label{Eq_level}
\ee
for long times.

Therefore in real time the walker is logarithmically slow with $R(t)
\sim \log(t)^{2/3}$ in one dimension and $R(t) \sim \log(t)^{1/2}$ in
higher dimensions. The logarithmic dependence of the RMS distance of
the walker is just the consequence of Eq. \ref{Eq_level} as a result
of which, the probability of making a jump to a neighbouring site
decreases as $1/t$. It is thus valid in any dimension. The value of
the exponent of the log in one dimension, is however a non-trivial
consequence of the coupling between the walker and the medium which
induces long-range memory effects. 

\section{Conclusion}
In summary, we have introduced and studied a simple model of a walker
interacting with its environment. By choosing the correct measures for
describing the time and the potential, 
we could map the SQW problem to the TSAW
model and thus use the known exact results in the latter case,
to describe the motion of the SQW walker. In
addition, we have also studied the emerging landscape. Though the
rules for the SQW are entirely 
local, a relationship with the so called extremal models could be
established. The critical dimension is $d_c=2$ below which the
potential landscape gets self-affine and the walker super-diffusive in
terms of moves. In real time the walker is logarithmically slow in any
dimension. 

{\bf Acknowledgment:} 
This work was partially supported by OTKA T029985.


\begin{thebibliography}{999}

\bibitem{Alexander} S. Alexander, J. Bernasconi, W. R. Schneider and
R. Orback, Rev. Mod. Phys. {\bf 53}, (1981) 175.\\
S. Havlin, D. ben-Avraham, Adv. Phys. {\bf 36}, (1987) 695.\\
J.-P. Bouchaud and A. Georges, Phys. Rep. {\bf 195}, (1990) 127.\\
J.-P. Bouchaud, A. Comtet, A. Georges and P. Le Doussal, Ann. Phys.
{\bf 201}, (1990) 285.

\bibitem{Sinai} Ya G. Sinai, Theory Probab. Appl. {\bf 27}, (1982) 247.

\bibitem{Fisher} P. Le Doussal, C. Monthus and D. S. Fisher, Phys.
Rev. E {\bf 59}, (1999) 4795.

\bibitem{Bernasconi} J. Bernasconi, H. Beyeler, S. Str\"assler and
S. Alexander, Phys. Rev. Lett. {\bf 42}, (1979) 819.

\bibitem{monthus} C. Monthus and J.P. Bouchaud, J. Phys. A: Math. \& Gen.
{\bf 29}, (1996) 3847.

\bibitem{Eulwalk} V. B. Priezzhev, D. Dhar, A. Dhar and S.
Krishnamurthy, Phys. Rev. Lett. {\bf 77}, (1996) 5079.

\bibitem{Bak} P. Bak, C. Tang and K. Weisenfeld, Phys. Rev. Lett. {\bf
59}, (1987) 381.

\bibitem{ddhar} D. Dhar, Phys. Rev. Lett. {\bf 64}, (1990) 1613.

\bibitem{activ_w} D. Helbing, F. Schweitzer, J. Keltsch, P. Moln\'ar,
Phys. Rev. E {\bf 56}, (1997) 2527, and references therein.

\bibitem{Czirok} E. Ben-Jacob, O. Shochet, A. Tenenbaum, I. Cohen, A.
Czir\'ok and T. Vicsek, Fractals {\bf 2}, (1994) 15.

\bibitem{deGennes} P. G. de Gennes, {\it Scaling Concepts in Polymer
Physics}, (Cornell University Press, 1979).
 
\bibitem{TSAWorig} D. J. Amit, G. Parisi and L. Peliti, Phys. Rev. E
{\bf 27}, (1983) 1635.

\bibitem{TKKR} J. T\"or\"ok, S. Krishnamurthy, J. Kert\'esz, S. Roux,
Phys. Rev. Lett. {\bf 84}, (2000) 3851.



\bibitem{JCAdA} R. Rammal, J-C. Angles d'Auriac and A. Benott, J. Phys.
A: Math. \& Gen. {\bf 17}, (1984) L9.

\bibitem{BerPie} J. Bernasconi and L. Pietronero, Phys. Rev. B {\bf
29}, (1984) 5196.

\bibitem{FamDaoud} F. Family and M. Daoud, Phys. Rev. B {\bf 29},
(1984) 1506.

\bibitem{Pietronero} L. Pietronero, Phys. Rev. B {\bf 27}, (1983) 5887.

\bibitem{BToth} B. T\'oth, J. Stat. Phys. {\bf 77}, (1994) 17\\
B. T\'oth, Ann. Probab. {\bf 23}, (1995) 1523.

\bibitem{FS} F. Family and T. Vicsek, {\it Dynamics of fractal
surfaces}, (Singapore, World Scientific 1991).

\bibitem{supri} S. Krishnamurthy, A. Tanguy and S. Roux, Eur. Phys.
J. B {\bf 15}, (2000) 149.


\bibitem{Maslov} M. Paczuski, S. Maslov and P. Bak, Phys. Rev. E {\bf
53}, (1995) 414.

\bibitem{Tang} A. Tanguy, M.Gounelle and S. Roux, Phys. Rev. E {\bf
58}, (1998) 1577.

\end{thebibliography}
\end{document}